\documentclass[%
reprint,
showpacs,preprintnumbers,
 amsmath,amssymb,
 aps,
]{revtex4-1}

\usepackage{graphicx}
\usepackage{dcolumn}
\usepackage{bm}
\usepackage{subfigure} 
\usepackage{color,soul}
\usepackage{ulem}

\begin{document}


\title{Spreading law of non-Newtonian power-law liquids on a spherical substrate by an energy-balance approach
}

\author{Masao Iwamatsu}
\email{iwamatsu@ph.ns.tcu.ac.jp}
\affiliation{%
Department of Physics, Faculty of Liberal Arts and Sciences, Tokyo City University, Setagaya-ku, Tokyo 158-8557, Japan
}%



\date{\today}

\begin{abstract}
The spreading of a cap-shaped spherical droplet of non-Newtonian power-law liquids, both shear-thickening and shear-thinning liquids, that completely wet a spherical substrate is theoretically investigated in the capillary-controlled spreading regime.  The crater-shaped droplet model with the wedge-shaped meniscus near the three-phase contact line is used to calculate the viscous dissipation near the contact line.  Then the energy balance approach is adopted to derive the equation that governs the evolution of the contact line.  The time evolution of the dynamic contact angle $\theta$ of a droplet obeys a power law $\theta \sim t^{-\alpha}$ with the spreading exponent $\alpha$, which is different from Tanner's law for Newtonian liquids and those for non-Newtonian liquids on a flat substrate.  Furthermore, the line-tension dominated spreading, which could be realized on a spherical substrate for late-stage of spreading when the contact angle becomes low and the curvature of the contact line becomes large, is also investigated. 
\end{abstract}

\pacs{64.60.Q-}
\keywords{Spreading, Spherical Substrate, Energy balance}
\maketitle

\section{Introduction}
The spreading of a liquid droplet on a solid substrate plays fundamental roles in many natural phenomena and industrial applications~\cite{deGennes1985,Daniel2006,Bonn2009}.  In particular, the spreading of non-Newtonian liquids is an important industrial process in printing, painting, coating, and various manufacturing processing because numerous polymer solutions and particulate suspensions exhibit non-Newtonian behaviors~\cite{Liang2009}.  Even though, the spreading of a liquid droplet on a solid substrate is a complicated phenomena where many factors come into play, the time evolution of the spreading of a Newtonian liquid droplet on a flat solid surface can be usually described by simple universal power laws~\cite{Hoffman1975,Voinov1976,Tanner1979,Hervet1984,deGennes1985,Seaver1994,deRuijter2000,Daniel2006}.  The most well-known law called Tanner's law describes the spreading of a small non-volatile droplet of Newtonian liquids on a completely wettable flat substrate.  This law was derived theoretically using several different approaches~\cite{Voinov1976,Tanner1979,Hervet1984} and confirmed experimentally~\cite{Tanner1979,deRuijter1999,Rafai2004}. However, most of the theoretical as well as experimental work was confined to a droplet of Newtonian liquids on a flat substrate.  Furthermore, the line-tension effect, which can be important on a spherical substrate when the contact angle becomes low, has not been considered except for a few theoretical works on a flat substrate~\cite{Fan2006,Mechkov2009}.

In a series of our previous works~\cite{Iwamatsu2015,Iwamatsu2016a,Iwamatsu2016b}, we investigated the wetting of a spherical substrate by a spherical cap-shaped droplet.  We showed that the wetting of a spherical substrate was totally different from that of a flat substrate, in particular, when the line tension was important.  For example, the complete wetting state can be realized by {\it positive} line tension on a spherical substrate~\cite{Iwamatsu2016a}, while it can be realized by {\it negative} line tension on a flat substrate~\cite{Widom1995}. Although the magnitude of the line tension is believed to be small~\cite{Pompe2000,Wang2001,Checco2003,Schimmele2007,Berg2010} so that the size of the droplet must be nano-scale, there is some argument that the line tension could be a few order of magnitude larger~\cite{Herminghaus2006, Law2017} than it has bee predicted so far when the gravitation can be important. Furthermore, the effect of line tension will be enhanced on a spherical substrate when the complete-wetting state is approached because the radius of the contact line vanishes and, therefore, the curvature of the contact line diverges.

\begin{figure}[htbp]
\begin{center}
\includegraphics[width=0.80\linewidth]{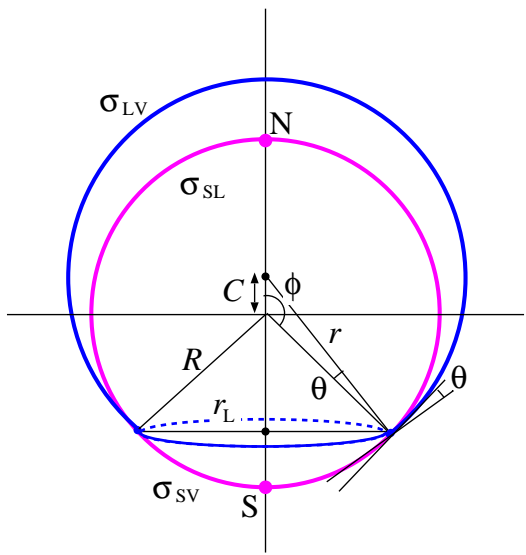}
\caption{
A spherical liquid droplet spreading from the north pole {\it N} on a convex spherical substrate towards the south pole {\it S}.  The center of the droplet with radius $r$ and that of the spherical substrate with radius $R$ are separated by a distance $C$.  The radius of the contact line is denoted by $r_{\rm L}$ and the dynamic contact angle is denoted by $\theta$, which is related to the half of the central angle $\phi$.  In the complete-wetting limit $\theta\rightarrow 0^{\circ}$ ($\phi\rightarrow 180^{\circ}$), the contact line shrinks and approaches the south pole {\it S} of the spherical substrate.  
}
\label{fig:NN1}
\end{center}
\end{figure}

In the present study, we will extend our previous study~\cite{Iwamatsu2017} of the spreading of a Newtonian-liquid droplet on a spherical substrate.  We will consider the problem of spreading of a cap-shaped spherical droplet of non-volatile non-Newtonian liquids gently placed on the north pole {\it N} of a spherical substrate and spreading towards the south pole {\it S} (Fig.~\ref{fig:NN1}) using the energy-balance approach~\cite{Hervet1984,deGennes1985,Daniel2006}, which can easily include the line-tension effect~\cite{Mechkov2009}.  The size of the droplet is assumed to be smaller than the capillary length so that the cap-shaped spherical meniscus is justified~\cite{deGennes1985,Bonn2009}.  Although, several theoretical as well as experimental works on the spreading of non-Newtonian liquids have already appeared~\cite{Rafai2004,Carre2000,Starov2003,Betelu2004,Wang2007,Liang2009,Dandapat2010,Liang2012}, they consider the spreading only on a flat substrate.  Furthermore, the effect of line tension has not been included.  This paper, together with our previous paper~\cite{Iwamatsu2017}, will be the first step towards the detailed understanding of the spreading of a droplet on a spherical substrate.  This paper will, hopefully, encourage more researchers and developers to pay attention to the interesting problem of spreading on spherical substrates~\cite{Tao2011,Eral2011,Extrand2012}.

\section{\label{sec:sec2}Spreading law of non-Newtonian liquids on a spherical substrate}

We will consider the spreading dynamics of a cap-shaped droplet of non-Newtonian power-law liquids~\cite{Liang2009} on a spherical substrate (Fig.~\ref{fig:NN1}).  The apparent viscosity $\mu$ depends on the shear rate $\dot{\gamma}$ through
\begin{equation}
\mu = \kappa \dot{\gamma}^{n-1},
\label{eq:N1}
\end{equation}
where $\kappa$ is a consistency coefficient~\cite{Wang2007,Dandapat2010,Liang2012}. The power exponent $n$ characterizes the non-Newtonian liquids.  When $n>1$, the liquid is called shear thickening.  When $n<1$, it is called shear thinning.  The Newtonian liquids correspond to $n=1$.

\begin{figure}[htbp]
\begin{center}
\includegraphics[width=0.8\linewidth]{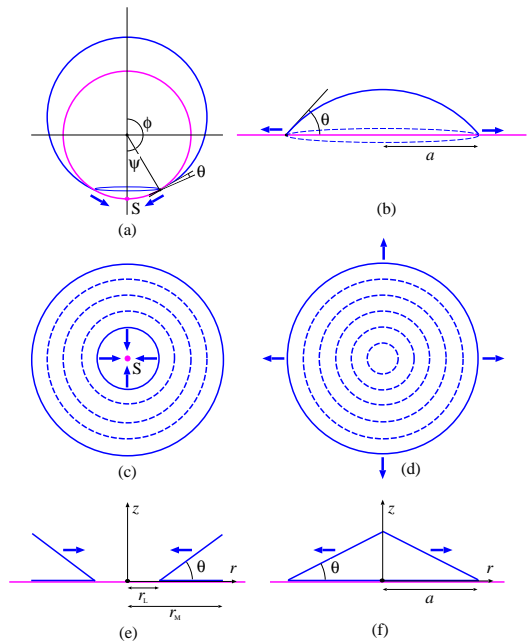}
\caption{
(a) A spreading droplet on a spherical substrate toward the complete-wetting state.  The three-phase contact line shrinks towards the south pole {\it S} of the spherical substrate.  (b) A spherical droplet on a flat substrate.  In this case, the three-phase contact line expands towards infinity.  Therefore, the line tension on a spherical substrate plays opposite role to that on a flat substrate.   (c) The bottom view of the spreading droplet, whose contact line shrinks towards south pole {\it S} of a spherical substrate, and (d) the top view of the spreading droplet whose contact line expands towards infinity on a flat substrate.  (e) The crater-shaped model of spreading droplet on a spherical substrate, and (f) the cone-shaped model of droplet on a flat substrate.  The meniscus is approximated by the wedge in both cases.
 }
\label{fig:NN2}
\end{center}
\end{figure}

To study the spreading on a spherical substrate, we will concentrate on the late-stage of the spreading shown in Fig.~\ref{fig:NN2}(a).  We will model the spreading of the three-phase contact line towards the south pole {\it S} of the spherical substrate as the shrinking circular contact line towards the singular point of two-dimensional flat surface as shown in Fig.~\ref{fig:NN2}(c).  Therefore, the spreading droplet on a spherical substrate is now modeled by a shrinking crater on a flat substrate [Fig.~\ref{fig:NN2}(e)] in contrast to the spreading droplet on a flat substrate, which is frequently modeled by a spreading cone [Fig.~\ref{fig:NN2}(b), (d), and (f)].

In the vicinity of the contact line, the liquid flow field can be described by Navier-Stokes equation of axial symmetry~\cite{Carre2000,Wang2007,Liang2012}
\begin{equation}
\frac{\partial P}{\partial r} = \frac{\partial}{\partial z}\left(\mu \frac{\partial u}{\partial z}\right),
\label{eq:N2}
\end{equation}
where $P$ is the pressure and $u$ is the velocity field of the liquid.  Therefore, the shear rate in Eq.~(\ref{eq:N1}) is given by
\begin{equation}
\dot{\gamma}=\frac{\partial u}{\partial z}.
\label{eq:N3}
\end{equation}

Equation (\ref{eq:N2}) with Eqs.~(\ref{eq:N1}) and (\ref{eq:N3}) should be solved with the boundary conditions of no shear at the free liquid-vapor interface and no slip at the solid-liquid interface given by
\begin{eqnarray}
\frac{\partial u}{\partial z} &=& 0,\;\;\;\;z=h\left(r\right),
\label{eq:N4} \\
u &=& 0,\;\;\;\; z=0,
\label{eq:N5}
\end{eqnarray}
where $h=h\left(r\right)$ is the height of the liquid-vapor surface from the substrate at the distance $r$ from the center of the droplet [Fig.~\ref{fig:NN2}(e)].  Integrating Eq.~(\ref{eq:N2}) twice with the boundary conditions (\ref{eq:N4}) and (\ref{eq:N5}), we obtain the flow field
\begin{equation}
u=\frac{n}{n+1}\kappa^{-\frac{1}{n}}\left(\frac{\partial P}{\partial r}\right)^{\frac{1}{n}}\left[\left(z-h\right)^{\frac{n+1}{n}}-\left(-h\right)^{\frac{n+1}{n}}\right],
\label{eq:N6}
\end{equation}
The average flow rate $U$ is given by
\begin{equation}
U=\frac{1}{h}\int_{0}^{h} u dz = -\frac{n}{2n+1}\left(\frac{\partial P}{\partial r}\right)^{\frac{1}{n}}\left(-h\right)^{\frac{n+1}{n}},
\label{eq:N7}
\end{equation}
which is identified with the spreading velocity of the three-phase contact line~\cite{deGennes1985}.  Then, the flow field in Eq.~(\ref{eq:N6}) is written as
\begin{equation}
u=\frac{2n+1}{n+1}U\left[1-\left(1-\frac{z}{h}\right)^{\frac{n+1}{n}}\right].
\label{eq:N8}
\end{equation}
Therefore, the viscous dissipation near the axial-symmetric contact line [Fig.~\ref{fig:NN2}(e)] will be calculated from 
\begin{eqnarray}
\dot{\Sigma}_{\rm drop}&=&\int_{V} \mu\left(\frac{du}{dz}\right)^{2}dV
\nonumber \\
&=&\int_{0}^{2\pi}\int_{r_{\rm L}+\Delta r}^{r_{\rm M}}\int_{0}^{h} \mu\left(\frac{du}{dz}\right)^{2}
dz rdr d\varphi
\label{eq:N9}
\end{eqnarray}
which can be integrated to give
\begin{equation}
\dot{\Sigma}_{\rm drop}=2\pi\kappa\left(\frac{2n+1}{n}\right)^{n}U^{n+1}\int_{r_{\rm L}+\Delta r}^{r_{\rm M}}\frac{1}{h^{n}}rdr,
\label{eq:N10}
\end{equation}
where we introduce the upper bound $r_{\rm M}$ and the cutoff $\Delta r$ to the lower bound $r_{\rm L}$ at the contact line to avoid the singularity. 

Assuming an axial-symmetric wedge-shaped meniscus of the crater model shown in Fig.~\ref{fig:NN2}(e) given by
\begin{equation}
h\left(r\right)=\theta\left(r-r_{\rm L}\right),\;\;\;r_{\rm M}>r>r_{\rm L},
\label{eq:N11}
\end{equation}
Eq,~(\ref{eq:N10}) can be integrated to give
\begin{equation}
\dot{\Sigma}_{\rm drop}=2\pi\lambda\left(\frac{2n+1}{n}\right)^{n}\frac{\kappa U^{n+1}}{\theta^{n}}r_{\rm L}^{2-n},
\label{eq:N12}
\end{equation}
where
\begin{eqnarray}
\lambda&=&\frac{\left(\Lambda-1\right)^{1-n}}{\left(n-1\right)\left(n-2\right)}
-\frac{\left(\Lambda-1\right)^{1-n}\Lambda}{\left(n-2\right)}
+\frac{\delta^{1-n}}{\left(n-1\right)}+\frac{\delta^{2-n}}{\left(n-2\right)},
\nonumber \\
&&\;\;\;\;\left(n\neq 1\right)
\label{eq:N13}
\end{eqnarray}
when $n\neq 1$ (non-Newtonian liquid), where $\Lambda=r_{\rm M}/r_{\rm L}\gg 1$ and $\delta=\Delta r/r_{\rm L}<1$.
For shear-thickening liquid ($n>1$), the singularity
\begin{equation}
\lambda \sim \delta^{1-n}
\label{eq:N14}
\end{equation}
occurs as $\delta\rightarrow 0$. However, this singularity will not be important as $r_{\rm L}\rightarrow 0$ and, therefore, $\delta\rightarrow 0$ will not be realized, for the spreading on a spherical substrate in contrast to the spreading on a flat substrate where $r_{\rm L}\rightarrow\infty$.

For a Newtonian liquid with $n=1$, we have
\begin{equation}
\lambda=\Lambda+\ln\left(\Lambda-1\right)-1-\delta-\ln\delta,
\label{eq:N15}
\end{equation}
which shows the well-known singularity
\begin{equation}
\lambda\sim\ln\delta.
\label{eq:N16}
\end{equation}
Again, this singularity will not be important for the spreading on a spherical substrate as the contact line will shrink and $r_{\rm L}\rightarrow 0$.

The thermodynamic driving force (capillary force) $f_{\rm L}$ per unit length acting at the three phase contact line is given by~\cite{Iwamatsu2017}
\begin{equation}
f_{\rm L}=\sigma_{\rm LV}\left(\cos\theta_{\rm Y}-\cos\theta\right) - \frac{\tau}{R\tan\phi}
\label{eq:N17}
\end{equation}
where $\sigma_{\rm LV}$ is the liquid-vapor surface tension and $\phi$ is half of the central angle (Fig.~\ref{fig:NN1}), which is related to the dynamic contact angle $\theta$ through
\begin{eqnarray}
C\cos\phi &=& R-r\cos\theta,
\nonumber \\
C\sin\phi &=& r\sin\theta,
\label{eq:N18}
\end{eqnarray}
where $C$ is the distance of two spheres with radius $r$ and $R$ (Fig.~\ref{fig:NN1}).  Therefore
\begin{equation}
\tan\phi=\frac{r\sin\theta}{R-r\cos\theta}.
\label{eq:N19}
\end{equation}
The energy-balance condition at the contact line with radius $r_{\rm L}$
\begin{equation}
2\pi r_{\rm L} f_{\rm L}U=\dot{\Sigma}_{\rm drop}
\label{eq:N20}
\end{equation}
is given by
\begin{equation}
\theta^{n}\left[\left(\cos\theta_{\rm Y}-\cos\theta\right)-\frac{\tilde{\tau}}{\tan\phi}\right]=\lambda\left(\frac{2n+1}{n}\right)^{n}r_{\rm L}^{1-n}\frac{\kappa U^{n}}{\sigma_{\rm LV}},
\label{eq:N21}
\end{equation}
where
\begin{equation}
\tilde{\tau}=\frac{\tau}{\sigma_{\rm LV}R}
\label{eq:N22}
\end{equation}
is the scaled line tension relative to the liquid-vapor surface tension $\sigma_{\rm LV}$. 
Equation (\ref{eq:N21}) can be applied both to the complete wetting ($\theta_{\rm Y}=0^{\circ}$) and incomplete wetting ($\theta_{\rm Y}\neq 0^{\circ}$).  This energy-balance approach in Eq.~(\ref{eq:N20}) is valid only when the spreading velocity $U$ is low and the viscous length scale is larger than other length scales~\cite{Bonn2009} so that the inertial effect can be neglected.  In the inertial regime when the velocity is high, the dissipation can be neglected and the capillary energy is directly transformed into kinetic energy~\cite{deGennes1985}.

Now we will consider $\theta\ll 1$ on a hydrophilic substrate with $\theta_{\rm Y}\ll 1$.  Then, the angle $\psi$ defined in Fig.~\ref{fig:NN2}(a) becomes
\begin{equation}
\psi=\pi-\phi \rightarrow 0,
\label{eq:N23}
\end{equation}
then
\begin{equation}
\tan\phi\simeq -\psi \rightarrow -\frac{r_{0}}{r_{0}-R}\theta
\label{eq:N24}
\end{equation}
from Eq.~(\ref{eq:N19}) where $r_{0}$ is the radius of the droplet when it completely wets and encloses the spherical substrate of radius $R$, which can be specified by the droplet volume $V_{0}$ through
\begin{equation}
V_{0}=\frac{4\pi}{3}\left(r_{0}^{3}-R^{3}\right)
\label{eq:N25}
\end{equation}
and the radius $r_{\rm L}$ of the contact line becomes
\begin{equation}
r_{\rm L}=R\sin\phi\simeq R\psi\simeq \frac{r_{0}}{r_{0}-R}R\theta.
\label{eq:N26}
\end{equation}
Then, the energy-balance condition in Eq.~(\ref{eq:N21}) can be written as
\begin{eqnarray}
\theta^{2n-1}\left[\frac{1}{2}\left(\theta^{2}-\theta_{\rm Y}^{2}\right)+\frac{r_{0}-R}{r_{0}\theta}\tilde{\tau}\right]&& \nonumber \\
=\lambda\left(\frac{2n+1}{n}\right)^{n}\left(\frac{r_{0}}{r_{0}-R}R\right)^{1-n}\frac{\kappa U^{n}}{\sigma_{\rm LV}}.
\label{eq:N27}
\end{eqnarray}
This relationship between the dynamic contact angle $\theta$ and the spreading velocity $U$ given by $\theta^{2n+1}\propto U^{n}$ for the complete wetting ($\theta_{\rm Y}=0^{\circ}$) without line tension ($\tilde{\tau}=0$) is very similar to those derived for a droplet on a flat substrate~\cite{Liang2012,Wang2007,Betelu2004,Starov2003,Carre2000}, though the exponent is different.  For example, Carr\'e and Dustache~\cite{Carre2000} and Wang {\it et al.}~\cite{Wang2007} derived $\theta^{n+2}\propto U^{n}$ using the two-dimensional wedge model.  Liang {\it et al.} derived $\theta^{\left(2n+7\right)/3}\propto U^{n}$ using the three dimensional cone-shaped model~\cite{Liang2012} similar to our three-dimensional crater-shaped model.  The difference of exponent comes from the difference of geometry of the cone-shaped droplets [Fig.~\ref{fig:NN2}(f)] on a flat substrate from that in crater-shaped droplets [Fig.~\ref{fig:NN2}(e)] on a spherical substrate.  All those results, including our Eq.~(\ref{eq:N27}) reduce to the universal law~\cite{Hoffman1975,Voinov1976,Tanner1979,Seaver1994}
\begin{equation}
\theta^{3} \propto {\rm Ca}
\label{eq:N28}
\end{equation}
for the complete-wetting ($\theta_{\rm Y}=0^{\circ}$) Newtonian fluids ($n=1$), where ${\rm Ca}=\kappa U/\sigma_{\rm LV}$ is the capillary number.

When the contact angle $\theta$ is low, the spreading speed $U$ on a spherical substrate is given by
\begin{equation}
U=\frac{d}{dt}R\phi=-R\dot{\psi}\simeq=-\frac{r_{0}}{r_{0}-R}R\dot{\theta}
\label{eq:N29}
\end{equation}
from Eq.~(\ref{eq:N24}), where $\dot{\theta}<0$.  For a completely wettable, hydrophilic substrate characterized by the Young's contact angle $\theta_{\rm Y}=0^{\circ}$, Eq.~(\ref{eq:N27}) is written as
\begin{equation}
\theta^{2n-1}\left(\frac{1}{2}\theta^{2}+\frac{r_{0}-R}{r_{0}\theta}\tilde{\tau}\right)=\left(-\Gamma\dot{\theta}\right)^{n}
\label{eq:N30}
\end{equation}
with
\begin{equation}
\Gamma=\left[\lambda\left(\frac{r_{0}}{r_{0}-R}\right)\frac{\kappa}{\sigma_{\rm LV}}\right]^{\frac{1}{n}}\left(\frac{2n+1}{n}\right)R^{\frac{1}{n}}.
\label{eq:N31}
\end{equation}
Note that this coefficient $\Gamma$ depends on the power exponent $n$, the radius of the substrate $R$ and the volume of the droplet $V_{0}$ through $r_{0}$, and, in particular, $\Gamma$ is proportional to the $1/n$-th power of the radius $R$ of the substrate $\Gamma\propto R^{1/n}$. When $n=1$, Eq.~(\ref{eq:N30}) reduces to the equation for Newtonian liquids~\cite{Iwamatsu2017}.

The time scale of spreading is characterized by $\Gamma$ given in Eq.~(\ref{eq:N31}).  Suppose the liquid is Newtonian polydimethylsiloxane (PDMS) with $n=1$, $\kappa=\mu=1.04$ Pa s, and $\sigma_{\rm LV}=21.2 {\rm mN}/{\rm m}$~\cite{Liang2012}, and the volume of the droplet is the same as that of the spherical substrate. Then, $r_{0}=2^{1/3}R\simeq 1.26R$ and the time scale $\Gamma$ is given by $\Gamma\simeq 7\times 10^{2}\lambda R$. The time scale is proportional to the radius $R$ of the substrate.  For example, the radius $R=1 {\rm mm}$ gives $\Gamma \simeq 1{\rm s}$ if $\lambda\simeq 1$. The time scale will be longer when the volume of the droplet is smaller, and will diverge as $r_{0}\rightarrow R$ from Eq.~(\ref{eq:N31}).

When the line tension can be neglected ($\tilde{\tau}=0$), we can solve Eq.~(\ref{eq:N30}) and obtain the time evolution of the contact angle
\begin{equation}
\theta = \theta_{0}\left[1+\theta_{0}^{\frac{n+1}{n}}\frac{n+1}{2^{\frac{1}{n}}n}\left(\frac{t}{\Gamma}\right)\right]^{-\frac{n}{n+1}},
\label{eq:N32}
\end{equation}
where $\theta_{0}$ is the contact angle at $t=0$.  Therefore, the time evolution of the contact angle is asymptotically given by
\begin{equation}
\theta \propto \left(\frac{t}{\Gamma}\right)^{-\frac{n}{n+1}},
\label{eq:N33}
\end{equation}
and the radius $r_{\rm L}$ of the contact circle shrinks according to
\begin{equation}
r_{\rm L} \propto \theta \propto \left(\frac{t}{\Gamma}\right)^{-\frac{n}{n+1}},
\label{eq:N34}
\end{equation}
whose spreading velocity decelerates according to
\begin{equation}
U \propto -\dot{\theta} \propto \left(\frac{t}{\Gamma}\right)^{-\frac{2n+1}{n+1}}
\label{eq:N35}
\end{equation}
from Eq.~(\ref{eq:N29}).  The characteristic time of evolution is $\Gamma\propto R$.  Therefore, the time scale is proportional to the size of spherical substrate. For Newtonian liquids with $n=1$, we can recover the results derived previously~\cite{Iwamatsu2017}.

On flat substrates, Liang {\it et al}.~\cite{Liang2012} derived the time evolution of the base radius $a$ [Fig.~\ref{fig:NN2}(b)] assuming the cone-shaped meniscus [Fig.~\ref{fig:NN2}(f)].  The time evolution is asymptotically given by
\begin{equation}
a \propto t^{\frac{n}{3n+7}},
\label{eq:N36}
\end{equation}
which reduces to the Tanner's law $a \propto t^{1/10}$ when $n=1$.  Since the droplet volume $V_{0}$ given by $V_{0}=\pi\theta a^{3}/4$ is fixed, Eq.~(\ref{eq:N36}) leads to the evolution law of the contact angle $\theta$ given by
\begin{equation}
\theta \propto t^{-\frac{3n}{3n+7}},
\label{eq:N37}
\end{equation}
which, again, leads to the result~\cite{deRuijter2000} for Newtonian liquids when $n=1$.  The spreading exponent $3n/\left(3n+7\right)$ on a flat substrate in Eq.~(\ref{eq:N37}) is different from the spreading exponent $n/\left(n+1\right)$ on a spherical substrate in Eq.~(\ref{eq:N33}).

\begin{figure}[htbp]
\begin{center}
\includegraphics[width=0.80\linewidth]{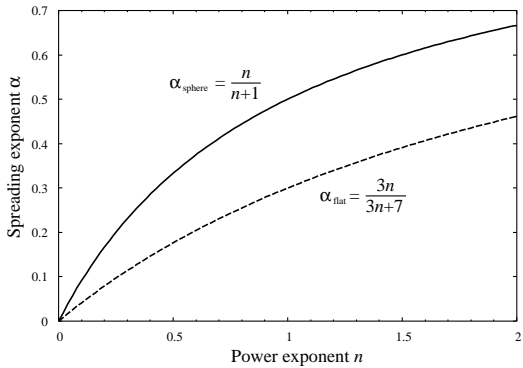}
\caption{
The spreading exponent $\alpha$ of the dynamic contact angle $\theta\propto t^{-\alpha}$ for spherical substrates (crater-shaped model) $\alpha_{\rm sphere}=n/\left(n+1\right)$ and for flat substrates (cone-shaped model) $\alpha_{\rm flat}=3n/\left(3n+7\right)$ as a function of the power exponent $n$ of viscous dissipation of power-law liquids. 
}
\label{fig:NN3}
\end{center}
\end{figure}

In Fig.~\ref{fig:NN3}, we compare the spreading exponent $\alpha$ of the dynamic contact angle
\begin{equation}
\theta \propto t^{-\alpha}
\label{eq:N38}
\end{equation}
for the spherical substrate $\alpha_{\rm sphere}=n/\left(n+1\right)$ and for the flat substrate $\alpha_{\rm flat}=3n/\left(3n+7\right)$.  Generally, $\alpha_{\rm sphere}>\alpha_{\rm flat}$ for the same non-Newtonian liquid characterized by the power exponent $n$.  The spreading on a spherical substrate is faster than that on a flat substrate, which can be easily understood from the efficiency of energy-dissipation.  The radius of contact line $r_{\rm L}$ shrinks on spherical substrates, while the radius expands infinity on flat substrates.  Therefore, energy dissipation by viscosity is less effective on flat substrates, which leads to weaker braking force.

When the contact angle $\theta$ becomes low, the line tension contribution in Eq.~(\ref{eq:N30}) could be important because the second term of the left-hand side will be dominant as $\theta\rightarrow 0$.  If the line tension is positive and dominant, the solution of Eq.~(\ref{eq:N30}) becomes
\begin{eqnarray}
\theta &=& \left[\frac{2-n}{n}\left(\frac{r_{0}-R}{r_{0}}\tilde{\tau}\right)^{\frac{1}{n}}\right]^{\frac{n}{2-n}}\left(\frac{t_{0}-t}{\Gamma}\right)^{\frac{n}{2-n}},
\nonumber \\
&&\;\;\;t<t_{0},\;\; n<2
\label{eq:N39}
\end{eqnarray}
when $n<2$, where $t_{0}$ is the time when the spreading will be completed ($\theta=0^{\circ}$), which is determined from the initial contact angle $\theta_{0}$ at $t=0$ and is given by
\begin{equation}
\frac{t_{0}}{\Gamma}=\theta_{0}^{\frac{2-n}{n}}\left(\frac{r_{0}-R}{r_{0}}\tilde{\tau}\right)^{-\frac{1}{n}}\frac{n}{2-n}.
\label{eq:N40}
\end{equation}
Therefore, the completion time $t_{0}$ depends strongly on the power exponent $n$.  It will be longer for shear thickening liquids with $n>1$ than for shear thinning liquids with $n<1$, and it will diverge as $n\rightarrow 2$.

On the other hand, the solution of Eq.~(\ref{eq:N30}), when $n>2$, is given by
\begin{equation}
\theta=\theta_{0}\left[1+\theta_{0}^{\frac{n-2}{n}}\frac{n-2}{n}\left(\frac{r_{0}-R}{r_{0}}\tilde{\tau}\right)^{\frac{1}{n}}\frac{t}{\Gamma}\right]^{-\frac{n}{n-2}},\;\;\; n>2
\label{eq:N41}
\end{equation}
where $\theta_{0}$ is the contact angle when $t=0$.  Then the evolution of the contact angle $\theta$ is given asymptotically by
\begin{eqnarray}
\theta &\propto&  \left(\frac{t_{0}-t}{\Gamma}\right)^{\frac{n}{2-n}},\;\;\;n<2,
\label{eq:N42} \\
&\propto& \left(\frac{t}{\Gamma}\right)^{-\frac{n}{n-2}},\;\;\;n>2.
\label{eq:N43}
\end{eqnarray}
The asymptotic form of the spreading velocity $U$ of the contact line is different for the shear-thinning liquid ($n<1$) to the shear thickening liquid ($n>1$) from Eq.~(\ref{eq:N30}).  The spreading velocity will be accelerated
\begin{equation}
U \propto -\dot{\theta} \propto \theta^{-\frac{2\left(1-n\right)}{n}}\rightarrow \infty
\label{eq:N44}
\end{equation}
as $\theta\rightarrow 0^{\circ}$ for shear-thinning liquids with $n<1$, and it will be decelerated
\begin{equation}
U \propto -\dot{\theta} \propto \theta^{\frac{2\left(n-1\right)}{n}}\rightarrow 0
\label{eq:N45}
\end{equation}
as $\theta\rightarrow 0^{\circ}$ for shear-thickening liquids with $n>1$.  The spreading velocity will be constant and the contact angle $\theta$ changes linearly with time as $\theta\propto t_{0}-t$ for Newtonian liquids with $n=1$~\cite{Iwamatsu2017}.  When the line tension is negative, the droplet cannot spread over the whole area of a spherical substrate. To achieve the complete wetting, a {\it positive} line tension is necessary on a spherical substrate~\cite{Iwamatsu2016a,Iwamatsu2017}.  In contrast, a {\it negative} line tension is necessary on a flat substrate~\cite{Mechkov2009,Iwamatsu2017}.

Our theoretical predictions must be checked by comparing them to experiments.  However, since the number of experimental works of spreading on a spherical substrate is very limited~\cite{Tao2011}, it is impossible to verify our theoretical predictions from experimental results at the present stage.  We must wait for the new experimental results.  Further experimental studies are certainly necessary.  

Our macroscopic model  known as the hydrodynamic model based on the viscous dissipation neglects various microscopic effects such as the friction at the contact line.  The energy dissipation due to the friction is consider in the so-called molecular-kinetic theory (MKT)~\cite{Blake1969,Bertrand2009}, which predicts scaling laws different from those of the hydrodynamic model~\cite{deRuijter2000}.  A mixed model, which takes into account both the friction and the viscous dissipation was used to analyze the spreading on incompletely wettable substrates~\cite{deRuijter2000}. It was found that the hydrodynamic model successfully describe the late stage of spreading, while MKT is appropriate to the early stage.  Hence, our model will be appropriate to the late stage of spreading on a spherical substrate as well. Furthermore, the friction should be more important at the edge of precursor film~\cite{Popescu2012}, which must exist ahead of the droplet on a completely wettable substrate.  Therefore, our hydrodynamic model is more appropriate to the spreading on a completely-wettable substrate.  Finally, it would be difficult to verify our theoretical predictions from the microscopic molecular-dynamics simulation, because the simulation time is limited to the early stage of spreading when MKT is more appropriate~\cite{Bertrand2009}.

There are several other microscopic effects such as the curvature dependence~\cite{Tolman1949} of the surface tension and the long-ranged liquid-substrate interaction called disjoining pressure~\cite{MacDowell2014}. In contrast to the droplet on a flat substrate [Fig.~\ref{fig:NN2}(b)], the shape of a droplet becomes almost spherical on a spherical substrate [Fig.~\ref{fig:NN2}(a)] in the late stage of spreading.  Therefore, the curvature of the liquid-vapor interface will be almost constant in time and its effect on the dynamics of spreading can be neglected.  The effect of disjoining pressure on the liquid-vapor surface tension will also be unimportant to the dynamics.  However, the disjoining pressure is directly responsible to the magnitude of line tension~\cite{Law2017}.  In fact, the line tension must depend on the radius of contact line~\cite{Napari2003};  otherwise, the line-tension contribution in Eq.~(\ref{eq:N17}) will diverge as the radius $r_{\rm L}$ vanishes ($\phi\rightarrow 180^{\circ}$).  Physically, the meniscus of the spreading droplet will merge to the flat precursor film~\cite{Popescu2012} as the complete-wetting state is approached, and the line tension must vanish. 

Seeing that the hydrodynamics model has already been fairly successful in describing the spreading of the droplet on a flat substrate~\cite{deRuijter2000,Carre2000,Liang2012}, our macroscopic model would also be appropriate to describe the late-stage of spreading.

\section{\label{sec:sec5} Conclusion}

In the present study, we consider the problem of spreading of a cap-shaped spherical droplet of non-Newtonian liquids on a spherical substrate using the energy balance approach.   The viscous dissipation is calculated using the crater-shaped model of droplet with the wedge-shaped meniscus.  We find scaling rules of the time evolution of the dynamic contact angle on a completely wettable spherical substrate, which are different from those for a droplet of non-Newtonian liquids spreading on a flat substrate~\cite{Carre2000,Liang2012}.  Since those scaling rules on a flat substrate are fairly successful in explaining the spreading of non-Newtonian liquids on a flat substrate~\cite{Carre2000,Liang2012},  experimental attempts to verify our scaling rule on a spherical substrate will be interesting.

In contrast to the spreading on a flat substrate where the three-phase contact line expands to infinity, the effect of line tension will be important in the late-stage of spreading on a spherical substrate where the radius of the contact line shrinks and the curvature diverges.  Furthermore, a {\it positive} line tension is necessary~\cite{Iwamatsu2016a,Iwamatsu2017} to realize complete wetting on a spherical substrate, while a {\it negative} line tension is necessary on a flat substrate~\cite{Mechkov2009}.  When the line tension is positive and dominant, the scaling rule for non-Newtonian liquids on a spherical substrate is different from that derived for Newtonian liquid on a spherical~\cite{Iwamatsu2017} as well as that on a flat substrate~\cite{Mechkov2009}.  Even though the magnitude of the line tension has been believed to be small, a gravitation assisted enhancement of the line tension~\cite{Herminghaus2006,Law2017} as well as the diverging contact-line curvature would make it possible to observe the line tension effect even in macroscopic droplets.

Finally, we notice that the spreading on a completely wettable spherical substrate is topologically different from that on a flat substrate.  On a spherical substrate,  the contact line shrinks, which involves a topological phase transition since the topology of the wetting film changes from a hollow to a spherical surface which encloses the spherical substrate.

\begin{acknowledgments}
This work was partially supported under a project for strategic advancement of research infrastructure for private universities, 2015-2020, operated by MEXT, Japan. A part of this work was done while the author was with the Department of Physics, Tokyo Metropolitan University as a visiting scientist.  The author is grateful to Professor Hiroyuki Mori and Professor Yutaka Okabe for continuous support and encouragement.  The author is also grateful to Professor Siegfried Dietrich (Max-Planck Institute for Intelligent Systems, Stuttgart) for sending him useful material on line tension.
\end{acknowledgments}




\end{document}